\documentclass[onecolumn]{elsart}
\usepackage{times}

\usepackage{graphicx}

\usepackage{amssymb}

\begin{document}

\begin{frontmatter}


\title{Resonating Valence Bond wave function: from lattice models to realistic
  systems} 

% Attention please ! do not leave blank lines here !
\author[ia1]{Michele Casula\corauthref{cor1}}
\corauth[cor1]{Corresponding Author:}
\ead{casula@sissa.it}
\author[ia1]{Seiji Yunoki}
\author[ia1]{Claudio Attaccalite}
\author[ia1]{Sandro Sorella}
\address[ia1]{International School for Advanced Studies (SISSA) Via Beirut
  2,4 34014 Trieste , Italy and INFM Democritos National Simulation Center,
  Trieste, Italy} 
%\address[ia2]{Address author2, Institution2}
% \thanks[label1]{On leave of absence from XXX }

\begin{abstract}
% Text of abstract
Although mean field theories have been very successful to predict
a wide range of properties for solids,
the discovery of high temperature superconductivity 
in cuprates supported the idea 
that strongly correlated materials cannot be qualitatively 
described by a mean field approach. After the original proposal 
by Anderson [P. W. Anderson, \emph{Science} {\bf 235}, 1196 (1987)],
there is now a large amount of numerical evidence that
the simple but general resonating valence bond (RVB) wave function
contains just those ingredients missing in uncorrelated theories, so that the
main features of electron correlation can be captured by the variational RVB
approach. 
Strongly correlated antiferromagnetic (AFM) systems, like $Cs_2  Cu  Cl_4$, 
displaying unconventional features of spin fractionalization, are also
understood within this variational scheme. 
From the computational point of view the remarkable feature of this 
approach is that several resonating valence bonds
can be dealt simultaneously with a single determinant,
at a computational cost growing with the number of electrons
similarly to more conventional methods, such as
Hartree-Fock  or Density Functional Theory.
Recently several molecules have been studied by using the RVB wave function;
we have always obtained total energies, bonding lengths and binding energies
comparable with more demanding multi configurational methods, 
and in some cases much better than single determinantal schemes.
Here we present the paradigmatic case of benzene.
\end{abstract}

\end{frontmatter}

% main text
\section{Introduction}
\label{intro} 
The variational approach, by providing 
an ansatz for  the ground state (GS) wave function of
a many body Hamiltonian, is one of the possible ways to analyze both
qualitatively and quantitatively a physical system. Moreover,
starting from the analytical properties of the variational wave function one 
is able in principle to understand and explain the mechanism underling a
physical phenomenon. For instance, the many body wave function of a quantum
chemical system can reveal the electronic structure of the compound and show
what is the nature of its chemical bonds. On the other hand, a very good
variational ansatz for a model Hamiltonian helps in predicting the ground
state properties and the qualitative picture of the system. In particular,
Pauling\cite{pauling} in 1949 introduced for the first time the concept of the
resonating valence bond (RVB) ansatz in order to describe the 
chemical structure of molecules such as  benzene and nitrous oxide; the idea
behind that concept is the superposition of all possible singlet pairs
configurations which link the various nuclear sites of a compound. He gave a
numerical estimate of the resonating energy in accordance with 
thermochemical data, showing the stability of the ansatz with respect to a
simple Hartree Fock valence bond approach. Few decades later, Anderson
\cite{anderson} in 1973
developed a mathematical description of the RVB wave function, in discussing
the ground state properties of a lattice frustrated model, i.e. the
triangular two dimensional Heisenberg antiferromagnet for spin $S=1/2$. His
first representation included an explicit sum over all the singlet pairs,
which turned out to be cumbersome in making quantitative calculations, the
number of configurations growing exponentially with the system size. 
Much later,   in
1987, with the aim to find an  explanation to high temperature (HTc)
 superconductivity by means of the variational approach, 
 he found a much more powerful 
 representation of the RVB state\cite{anderson2},  
based on the Gutzwiller projection $P$ of a BCS state
\begin{equation}
\label{pBCS}
P|\Psi\rangle = P~\Pi_\mathbf{k} (u_\mathbf{k} + v_\mathbf{k}
c^\dag_{\mathbf{k},\uparrow} c^\dag_{-\mathbf{k},\downarrow}) |0\rangle,
\end{equation}
which in real space and for a fixed number $N$ of electrons takes the form
\begin{equation}
\label{realpBCS}
P|\Psi\rangle = P~ \Sigma_{\mathbf{r},\mathbf{r}^\prime}
\left [ \phi(\mathbf{r}-\mathbf{r}^\prime) c^\dag_{\mathbf{r},\uparrow}
c^\dag_{\mathbf{r}^\prime,\downarrow} \right ]^{N/2} |0\rangle,
\end{equation}
where the \emph{pairing function} $\phi$ is the Fourier transform of
$v_\mathbf{k}/u_\mathbf{k}$. 
The Cooper pairs described by the BCS wave function are taken apart from each
other by the repulsive Gutzwiller projection, which avoids doubly occupied
sites; in this way the charge fluctuations present in the superconducting
ansatz are frozen and the system can become an insulator even 
when, according to band theory, it should be metallic.
%because there is one electron per site, 
%namely the free electron band is  half filled. 
The  wave function (\ref{realpBCS}) 
allows  a natural and simple description of a 
 superconducting state close to a Mott
insulator, opening the possibility  
for  a theoretical explanation 
of high temperature superconductivity, 
a phenomenon  discovered in 1986\cite{bednorz}, but not fully understood 
until now.    
Indeed, soon after this important experimental discovery, 
Anderson\cite{anderson2} suggested that the Copper-Oxygen  planes of cuprates
could be effectively described by an RVB state, and extensive developments
along this lines have subsequently taken place\cite{anderson3}. From the RVB
ansatz it is clear that the HTc superconductivity (SC) is essentially driven by
the Coulomb and magnetic interactions, with a marginal role played by phonons,
in spite of their crucial role in the standard BCS theory.
 As far as  the magnetic
properties are concerned,  the RVB state is  quite intriguing,
because  it represents an insulating 
phase of  an electron model with an odd number of electrons
 per unit cell, with vanishing  magnetic moment and without  
any finite  order parameter, 
namely a completely different picture 
from the  conventional mean field theory, where 
it is important to break the symmetry 
 in order to avoid the one electron per unit cell condition, 
incompatible with insulating behavior.
This rather unconventional RVB state is therefore called 
\emph{spin liquid}.
 
The structure of the paper is organized as follows: 
in section \ref{models} we present 
some numerical Monte Carlo studies of lattice models, where it is shown that,
once the Jastrow factor is included, the RVB wave
function is able to represent an exceptionally  good ansatz for the description
of the zero temperature properties of the systems studied, in very good  
agreement with the available experimental data. 
In section \ref{real}, we apply the same
variational wave function to quantum chemical systems, in particular to
benzene, where we exploit the Pauling's idea to study in a more systematic way
the role of the resonating valence bonds in this molecule, by performing
realistic \emph{ab initio} simulations. In the last section 
we make our final conclusions and highlight the perspectives of this
study.

\section{Lattice models}
\label{models} 
\begin{figure}[!ht]
%\vspace{5cm}
\includegraphics[width=13cm]{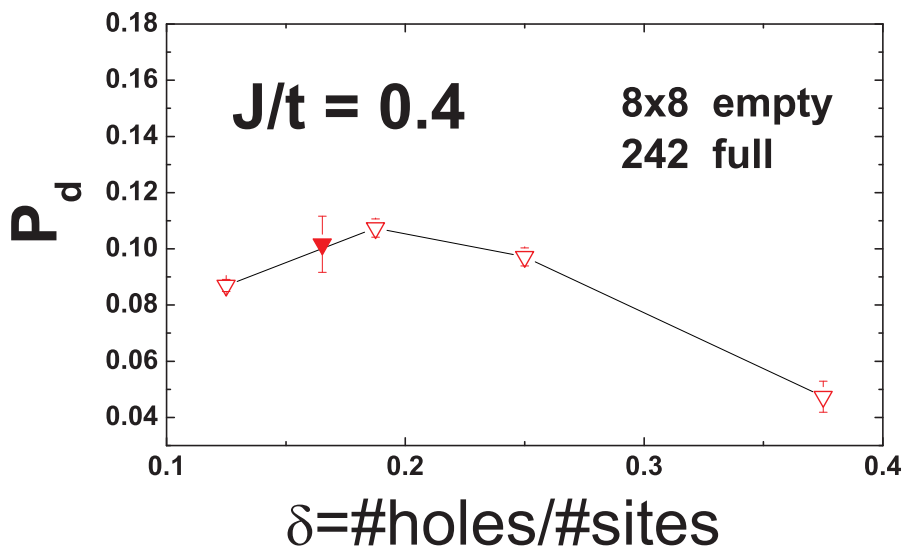}
%\vspace{5cm}
\caption{ Superconducting order parameter $P_d$ for the 2D $t-J$ model 
as a function of doping, at $J/t=0.4$, 
 calculated using variance extrapolation, using  
the projected BCS wave function defined in the text as an initial guess.
Data taken from Ref.~\cite{tj}} \label{figpd}
\end{figure}

In order to mimic in a simple way the essential features of a real
strongly correlated material, a lot of lattice models have been conceived so
far. One of the most important is the $t-J$ model, which takes into account
not only the charge degrees of freedom but also the magnetic superexchange
interactions:
\begin{equation}
  H = J \sum_{<i,j>} \left ( \mathbf{S}_i \cdot \mathbf{S}_j - \frac{1}{4} n_i
  n_j  \right ) - t \sum_{<i,j>,\sigma} \tilde{c}^\dag_{i,\sigma}
  \tilde{c}_{j,\sigma} + H.c.,
\end{equation}
where $\tilde{c}_{i,\sigma} = c_{i,\sigma} (1 - n_{i,\sigma})$, $\langle \ldots
 \rangle$ stands for nearest neighbor sites, and $n_i$ and $\textbf{S}_i$ are
 density and spin at site $i$, respectively. 
 In this case, the RVB wave function has shown to be an accurate
ansatz both for the chain, the two-leg ladder and the two dimensional (2D)
square lattice\cite{tj}, once a long range Jastrow factor has
been included besides the Gutzwiller projector. 
In particular for the 2D lattice, there is a rather clear evidence
that the GS of the doped model is superconducting, with an optimal doping
around $\delta \sim 0.18$; this  result has been obtained by performing Green
function Monte Carlo (GFMC) simulations within the fixed node (FN)
approximation up to 242 sites at various doping, and by calculating the
order parameter $P_d = 2 ~ \lim_{r \rightarrow \infty} \sqrt{|\Delta(r)|}$,
 where $\Delta(r)$ is the pair-pair correlation function.
If the state is a d-wave superconductor, 
$P_d$ must be non vanishing in the thermodynamic limit. 
At the variational level, the
RVB state gives a $P_d$ only 30 \% higher than the most accurate result
 calculated (see Fig.~\ref{figpd}); 
therefore the superconducting long range order is expected to remain 
stable against the
projection towards the GS of the system. Moreover, the RVB state is accurate
not only for SC but also for magnetic systems as well. Indeed, it is able to
capture both the quasi long range antiferromagnetic order of the $t-J$ chain
and the spin gapped behavior of the two-leg ladder. While the BCS part can
allow strong superconducting charge fluctuations, the Gutzwiller and
Jastrow parts control the charge correlation respectively at short and long
distance, allowing a quantitative description of the magnetic
behavior in low dimensional systems. 

One of the most non trivial questions which arise 
in a strongly correlated regime is
whether the ground state of a system containing \emph{only} repulsive
interaction can be superconductor. 
Surely such a state will not be found by
any mean-field theory, which needs an explicit or effective attractive
interaction in order to display a pairing strength among the
electrons. Instead this question can be addressed at least at the variational
level, dealing directly with strongly correlated variational wave functions
which may exhibit a superconducting behavior and be close to the
true GS of the system. Of course a clear indicator of the presence of
superconductivity is the SC order parameter $P_d$, but since its value is of
the same order as the quasiparticle weight, it can be too small to be detected
with a reasonable numerical precision. Therefore, with the aim of finding a good
probe for superconductivity, E. Plekhanov \emph{et al.}
\cite{plekhanov} defined a new suitable quantity $Z_c$, which measure the
\emph{pairing strength} between two electrons added to the GS wave function,
\begin{equation}
\label{zc}
Z_c = |F(\mathrm{shortest~ distance})| / \sqrt{\sum_{all~distances}
  F^2(R)} ,
\end{equation}
where $F(R)$ is related to the real space anomalous
part of the equal time Green function at zero temperature: 
\begin{equation}
F(R) = \langle N-2 | c_{i,\uparrow} c_{i+R,\downarrow} + c_{i+R,\uparrow}
c_{i,\downarrow} | N \rangle.
\end{equation}
For instance $Z_c=0$ for Fermi liquids, instead $Z_c \ne 0$ for
superconductors but also for non BCS systems which involve any kind of
pairing. The authors in Ref. \cite{plekhanov} applied this scheme
to the 2D Hubbard model
\begin{equation}
H = - t \sum_{<i,j>,\sigma} \tilde{c}^\dag_{i,\sigma}
  \tilde{c}_{j,\sigma} + h.c. + U \sum_i n_{i,\uparrow} n_{i,\downarrow} 
  - \mu N,
\end{equation}
where $\mu$ is the chemical potential and $N$ is the total number of
particles. Carrying out a projection Monte Carlo technique based on
auxiliary fields, they found that the GS of the undoped 2D Hubbard model
at half filling has a non vanishing pairing strength, although the system is
an insulator with antiferromagnetic long range order. This means that it is
not a band insulator, for which $Z_c$ should be zero, but an RVB Mott insulator
with a strong d--wave pairing character: 
indeed the RVB variational wave function is
very close to the projected GS, giving the same pairing strength and
a good variational energy. Moreover, the pairing strength decreases with
increasing doping, but it is still positive for the lightly doped Hubbard model,
suggesting that the system is ready to become superconductor, once the pairs
can condense and phase coherence can take place in the GS.

The accuracy of the RVB ansatz close to the Mott insulator transition (MIT) 
has been
pointed out also by Capello \emph{et al.} \cite{capello}, 
who undertake a variational Monte
Carlo study of the phases of the $t-t^\prime$ 1D Hubbard model with nearest
and next nearest neighbor hopping terms. The phase diagram of this model is
known from bosonization and density-matrix renormalization group calculations,
therefore it represents a good test case for the RVB variational wave function. 
When $t^\prime/t \lesssim 0.5$, the presence of a long range Jastrow factor
acting on a BCS state is a crucial ingredient to recover the insulator with
one electron per unit cell and without a broken translational
symmetry, i.e. an highly non trivial charge gapped state. 
On the other hand, once $t^\prime/t \gtrsim
0.5$ and $U/t$ is small, the \emph{same} wave function after optimization
is able to describe the
metallic state with strong superconducting fluctuations, 
namely a state with a finite spin gap.
The distinction between the metallic and the insulating state can be
made both by using the Berry phase\cite{resta}
 and by analyzing the behavior of the spin
and charge structure factor as $q \rightarrow 0$; in all cases,
the RVB state with an appropriate Jastrow factor reproduces very well the
known phases.  

\begin{figure}[!ht]
%\vspace{5cm}
\includegraphics[width=10cm,angle=-90]{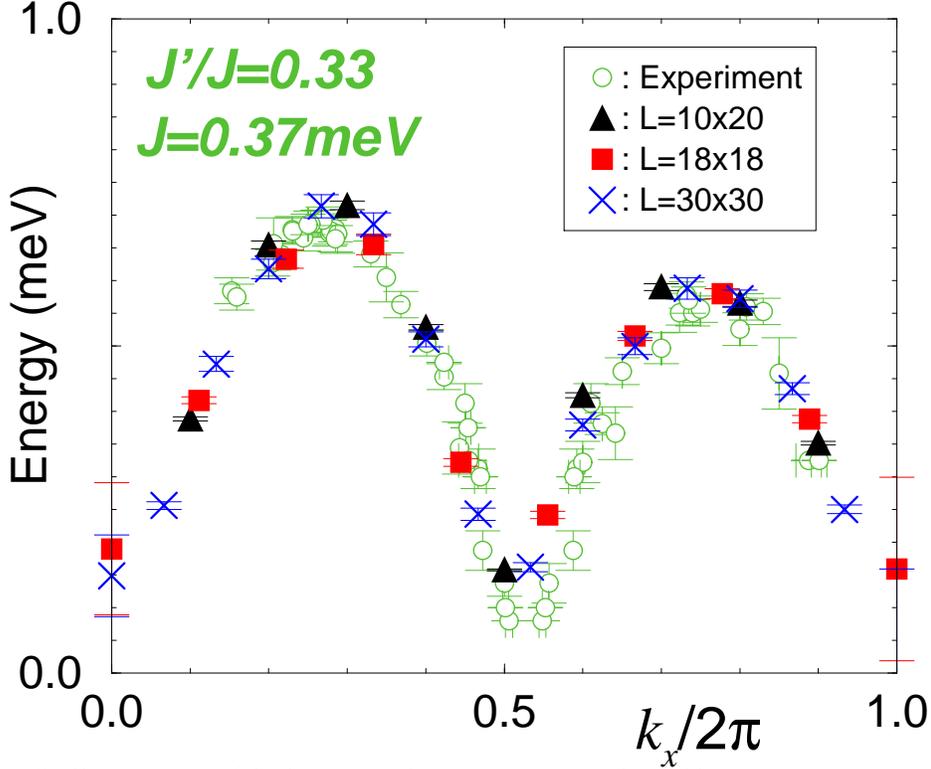}
%\vspace{5cm}
\caption{ 
Comparison of the lowest triplet excitations, evaluated by neutron 
scattering experiments on $Cs_2CuCl_4$ compound\protect\cite{coldearec}, 
with the QMC results, obtained using the lattice fixed node approximation 
and  the projected BCS state  to approximate  the signs  of the 
ground state wave function\protect\cite{yunoki}. There is no fitting 
parameter in the above comparison. 
} \label{figsqw}
\end{figure}

Not only the conducting properties of a strongly correlated model can be
reproduced by the RVB ansatz, but also the magnetic behavior. For instance,
in the case of the $t-t^\prime$ 1D Hubbard model, the variational wave
function drives the transition from the metallic to a dimerized insulator once
$U/t$ increases. For the 2D spin 1/2 AFM Heisenberg model on a
triangular lattice
\begin{equation}
H= J~\sum_{<i,j>} \mathbf{S}_i \cdot \mathbf{S}_j + J^\prime~\sum_{<<i,j>>}
\mathbf{S}_i \cdot \mathbf{S}_j , 
\end{equation}
with $J$ being the intra chain coupling and $J^\prime$ the inter chain one, 
the RVB wave function displays a stable spin liquid behavior, due to the
strong frustration of the system in the regime with $J'/J = 0.33$. Moreover
for $J=0.374 meV$, the model is able to represent a real system, the
$Cs_2CuCl_4$ compound
studied by Coldea \emph{et al.}\cite{coldea,coldearec}  who performed neutron
scattering experiments in order to determine the low lying magnetic
excitations. It turns out that the experimental data show an unconventional
behavior of the magnetic structure of the compound, with 
spin-1/2 fractionalized excitations and incommensurability. The numerical study
carried out in Ref.~\cite{yunoki} highlights that the
incommensurability comes from the frustration of the system and it is well
described by the RVB ansatz. The most impressive correspondence between the
experimental data and the numerical simulations is in the
spin-1 excitation spectrum (see Fig.\ref{figsqw}), 
obtained by GFMC calculations with an RVB state used as a guiding function.
As also shown in the same Fig.~\ref{figsqw}, it is evident that size effects 
are small and the comprairson between the numerical simulation and the 
experiment is particularly meaningful in this case.
This is possible within a Quantum Monte Carlo (QMC) scheme that allows to work 
with large enough systems sizes.

\section{Realistic systems}
\label{real}
As we have seen in the preceding section, the RVB wave
function can represent very well the GS of some strongly correlated
systems, which are described by a suitable lattice model, as in the case of 
$Cs_2CuCl_4$. Furthermore, following the seminal idea of Pauling, the
applicability of the RVB ansatz is not limited to the strongly correlated
regime close to the Mott transition or to spin frustrated models, but can be
extended to describe the electronic structure and the properties
of \emph{realistic} systems. Indeed the quantum chemistry community has
quite widely used the concept of pairing in order to develop a variational
wave function able to capture the most significant part of the electronic
correlation. Although only in 1987 Anderson discovered the link between the
explicit resonating valence bond representation and the projected-BCS wave
function, already in the 50' s Hurley \textit{et al.} \cite{hurley} introduced
the product of pairing functions as ansatz in quantum chemistry. 
Their wave function was called \emph{antisymmetrized
geminal power} (AGP) that has been shown to be the particle conserving
version of the BCS ansatz \cite{schrieffer}. It includes the single
determinantal wave function, i.e. the uncorrelated state, as a special case
and introduces correlation effects in a straightforward way, through the
expansion of the pairing function (in this context called geminal): 
therefore it was studied as a possible alternative to the other
multideterminantal approaches, but his success to describe correlation was
very much limited, because - we believe - the Jastrow term was not included.

For an unpolarized system containing $N$ electrons
(the first $N/2$ coordinates are referred to the up spin electrons)  
the AGP wave function is a $\frac{N}{2} \times \frac{N}{2}$ pairing matrix
determinant, which reads:  
\begin{equation}
\Psi_{AGP}(\textbf{r}_1,...,\textbf{r}_N) =
\det \left (\Phi_{AGP}(\textbf{r}_i,\textbf{r}_{j+N/2}) \right )~~{\rm for}~1\le i,j\le N/2, 
\end{equation}
and the geminal function is expanded over an atomic basis:
\begin{equation}
\Phi_{AGP}(\textbf{r}^\uparrow,\textbf{r}^\downarrow)
=\sum_{l,m,a,b}{\lambda^{l,m}_{a,b}\phi_{a,l}
(\textbf{r}^\uparrow)\phi_{b,m}(\textbf{r}^\downarrow)} ,  
\label{expgem}
\end{equation}
where indices $l,m$ span different orbitals centered on atoms $a,b$, and
$i$,$j$ are coordinates of spin up and down electrons respectively. 
It is possible to generalize the AGP many body wave function in order to deal
also with a polarized system. 
The geminal function may be viewed as an extension of the simple HF
wave function and in fact it
coincides with HF only when the number $M$ of non zero eigenvalues of the
$\lambda$ matrix is equal to $N/2$. 
It should be noticed that Eq.~\ref{expgem} is exactly the pairing function in
Eq.~\ref{realpBCS}, apart from the inhomogeneity of the former which reflects
the absence of the translational invariance of a generic molecular compound.
One of the main advantages of dealing with an AGP wave function is its
computational cost. Indeed one can prove that expanding the geminal by adding
more terms in the sum of Eq. \ref{expgem} is equivalent to introduce more
Slater determinants in the many body wave function, i.e. to have a
multireference total wave function, similar to those obtained in configuration
interaction (CI) or coupled cluster (CC) theories. 
But the computational cost of the AGP
ansatz still remains the same, since one needs to compute always just a
\emph{single} determinant.
% also for this type of multireference wave function.
This property  is expected to be important  for large scale simulations, since 
the number of determinants necessary for a satisfactory 
accuracy increases fast 
with the system size, limiting very much the 
applicability  of  CI and CC methods.

The simplest example which shows the essence of the AGP ansatz is the $H_2$
molecule. It is well known from textbooks that molecular orbital (MO) theory
at the HF level fails in predicting the binding energy and the bond length of
$H_2$, just because it overestimates the ionic terms contribution in the total
wave function if the antibonding molecular orbitals are not included. In spite
of this, the correct geminal expansion reads
\begin{equation}
\label{geminalh2}
\Phi_{AGP}(\textbf{r},\textbf{r}^\prime) = \lambda 
\phi^A_{1s}(\textbf{r})\phi^A_{1s}(\textbf{r}^\prime) 
+ \phi^A_{1s}(\textbf{r})\phi^B_{1s}(\textbf{r}^\prime)
+ A \leftrightarrow B,
\end{equation}  
where $\lambda$ can be tuned to regulate the weight of the different
resonating contributions and fulfill the size consistency when the two nuclei
are infinitely apart from each other ($\lambda \rightarrow 0$). 
Notice also that the chemical bond 
is represented in the geminal by a non vanishing value of $\lambda$
between the orbitals centered on the two different sites between which the
bond is formed.

Let us consider now a \emph{gas} of hydrogen dimers: in this case the geminal
will contain not only the terms in Eq.~\ref{geminalh2}, valid for just two
sites, but also the contributions from all the nuclei in the system. It is
clear that the AGP wave function will allow strong charge fluctuations 
around each $H$ pair, and therefore molecular sites with zero and
four electrons are permitted, leading to poor variational
energies. 
For this reason, 
the AGP alone is not sufficient, and it is necessary to introduce
a Gutzwiller-Jastrow factor in order to dump the expensive charge
fluctuations. Moreover only the AGP-Jastrow (AGP-J) wave function is the
real counterpart of the RVB ansatz of strongly correlated lattice models,
since the projection is essential to get the correct distribution of the
pairing in the compound. 
The AGP-J wave function has shown to be effective both in atomic
\cite{casula} and in molecular systems\cite{casula2}. 
Both the geminal and the Jastrow
play a crucial role in determining the remarkable accuracy of the many body
state: the former permits the correct treatment of the nondynamic correlation
effects, the latter allows the local conservation of charge in a complex
molecular system and also to fulfill the cusp conditions which 
make the geminal expansion rapidly converging to the lowest possible
variational energies.

 The study of the AGP-J variational ansatz 
with the inclusion of two and three body Jastrow factors
is possible by means of QMC techniques, which can deal 
explicitly with correlated wave functions. 
The optimization procedure, 
necessary to reach the lowest variational energy within the given
variational freedom, is feasible also in a stochastic Monte Carlo 
framework, after the recent developments in this field
(\cite{sorellaSR,filippi}). 

Benzene is the largest compound we have studied so far;
in order to represent its $^1 A_{1g}$ GS 
we have used a very simple one particle basis set: for the AGP, a 2s1p 
double zeta (DZ) Slater set centered  on the carbon atoms and a 1s 
single zeta (SZ) on the hydrogen.
For the 3-body Jastrow, a 1s1p DZ Gaussian set centered only on the
carbon sites has been chosen.
We started from a non
resonating 2-body Jastrow wave function, which dimerizes the ring and breaks
the full rotational symmetry, leading to  the Kekul\'e configuration.  
As we expected, the inclusion of the resonance between the two possible
Kekul\'e states lowers the variational Monte Carlo (VMC)
 energy by more than 2 eV. The wave function
is further improved by adding another type of resonance, that includes also the
Dewar contributions connecting third nearest neighbor carbons.
 As reported in Tab.~\ref{benzene}, the
gain with respect to the simplest Kekul\'e  wave function amounts to 4.2 eV, 
but the main improvement arises from the further 
inclusion of the three body Jastrow
factor, which allows to recover the $89 \%$ of the total atomization energy at
the VMC level. 
The main effect of the three body term is to 
keep  the total charge around the carbon sites  to approximately six electrons, 
thus penalizing   the double  occupation of the 
 $p_z$ orbitals. 

\begin{table}[!hbp]
\caption{\label{benzene}
Binding energies in $eV$ obtained by variational ($\Delta_{VMC}$) and diffusion
($\Delta_{DMC}$) Monte Carlo calculations with different trial wave functions
for benzene. In order to calculate the binding energies yielded 
by the two--body Jastrow, we used the atomic energies reported 
in Ref.~\cite{casula}.
The percentages ($\Delta_{VMC}(\%)$ and $\Delta_{DMC}(\%)$) 
of the total binding energies are also reported. Data are taken from
Ref.~\cite{casula2}.} 
%\begin{ruledtabular}
\begin{tabular}{l c c c c }
\hline
& \makebox[0pt][c]{$\Delta_{VMC}$} &
\makebox[0pt][c]{$\Delta_{VMC}(\%)$} & \makebox[0pt][c]{$\Delta_{DMC}$} &
\makebox[0pt][c]{$\Delta_{DMC}(\%)$} \\
\hline
\hline
Kekul\'e + 2body & -30.57(5) & 51.60(8)  & - & -  \\
resonating Kekul\'e + 2body & -32.78(5) & 55.33(8) & - & - \\
resonating Dewar Kekul\'e + 2body & -34.75(5) & 58.66(8) & -56.84(11) &
95.95(18) \\ 
Kekul\'e + 3body & -49.20(4) & 83.05(7) & -55.54(10) &  93.75(17) \\
resonating Kekul\'e + 3body & -51.33(4) & 86.65(7) & -57.25(9)  & 96.64(15)  \\
resonating Dewar Kekul\'e + 3body & -52.53(4) & 88.67(7)  &  -58.41(8) &
98.60(13) \\ 
full resonating + 3body & -52.65(4) & 88.869(7) & -58.30(8) &  
98.40(13) \\
\hline
\end{tabular}
%\end{ruledtabular}
\end{table}

\begin{figure}[!ht]
%\vspace{5cm}
\includegraphics[width=13cm]{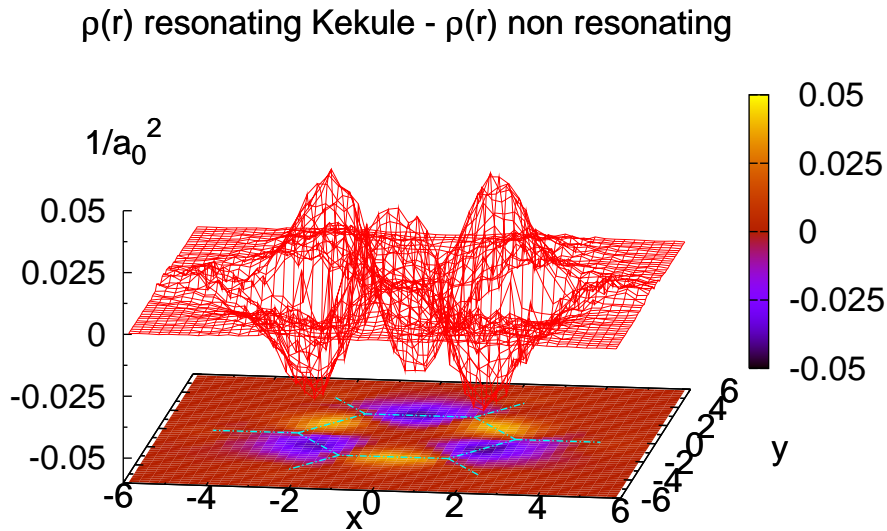}
%\vspace{5cm}
\caption{\label{density}
Electron density (atomic units) projected on the plane of $C_6H_6$.
The surface plot shows the difference between the resonating valence bond wave
function, with the correct $A_{1g}$ symmetry of the molecule, 
and a non-resonating one, which has the symmetry of the Hartree--Fock wave
function.}
\end{figure}

A more clear  behavior is found by
carrying out diffusion Monte Carlo (DMC) simulations: 
the interplay between the resonance among
different structures and the Gutzwiller-like correlation 
refines more and more the nodal surface topology, 
thus lowering the DMC energy by significant amounts. 
Therefore it is crucial to insert
into the variational wave function all these ingredients in order to have an
adequate description of the molecule. For instance, in
Fig. \ref{density} we report the density surface difference between the 
non-resonating 3-body Jastrow wave function, which breaks 
the $C_6$ rotational invariance, and the resonating Kekul\'e
structure, which preserves the correct $A_{1g}$ symmetry: the change in the
electronic structure is significant.
The best result for the binding energy  is obtained with the
Kekul\'e Dewar resonating 3 body wave function, which recovers the $98,6\%$ of
the total atomization energy with an absolute error of 0.84(8) eV.
As Pauling \cite{pauling} first pointed out, benzene is a genuine RVB system,
indeed it is well described by the AGP-J wave function.

\section{Conclusions}
\label{conclusions}
In this paper we have described a very powerful variational ansatz that 
has been  introduced to understand the properties of strongly correlated 
materials just after the discovery of HTc superconductivity. 
We have shown that the RVB wave function paradigm 
 is not only useful for describing the GS and low lying excitations 
of lattice models, such as Heisenberg or $t-J$ model, but is also suited for 
approaching  realistic systems, by considering explicitly the long 
range Coulomb repulsion and the full quantum mechanical interaction among  
electrons   within the Born-Oppenheimer approximation.
Moreover, by using the same type of wave function both for lattice model and 
realistic system, it is possible to have some insight in 
the electron correlation behind the latter and to check the reliability of
the model in predicting the properties of a real compund.
For instance the benzene molecule can be idealized by 
a six site ring Heisenberg model with one electron per site, 
in order to mimic the 
out of plane bonds of the real molecule, coming from the $p_z$ electrons
and leading to an antiferromagnetic superexchange interaction between 
nearest neighbor carbon sites.  
We have studied in this case the spin--spin correlations
\begin{equation}
C(i)=\langle  S^z_0  S^z_{i} \rangle,
\end{equation}
where the index $i$ labels consecutively the carbon sites starting 
from the reference $0$,
and the dimer--dimer correlations
\begin{eqnarray}
D(i) &=&  D_0(i)/C(1)^2-1, \nonumber \\
D_0(i) &=&  \langle  ( S^z_0 S^z_1) (S^z_i S^z_{i+1}) \rangle. 
\end{eqnarray}
Both correlation functions have to decay in an infinite ring, when
there is  neither magnetic ( $C(i) \to 0$ ), nor 
dimer ($D(i)\to 0$) long range order as in the true spin liquid ground state 
of the 1D Heisenberg infinite ring.  

\begin{figure}[!ht]
%\vspace{5cm}
\includegraphics[width=13cm]{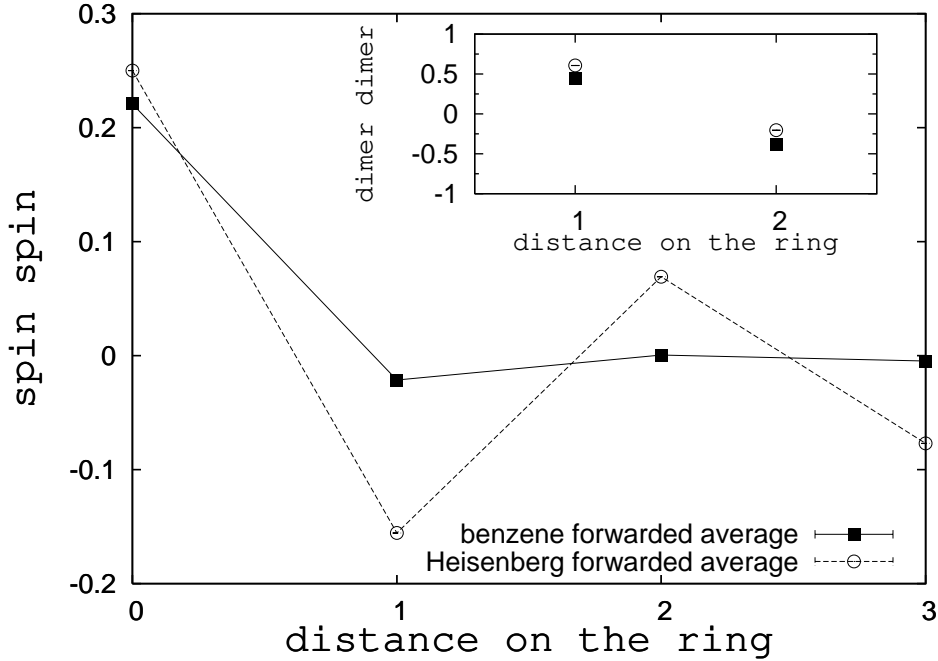}
%\vspace{5cm}
\caption{\label{spindimer}
Spin--spin correlation function for benzene (full squares) and for the
Heisenberg model (empty circles). In the inset, also the dimer--dimer
correlation function is reported with the same notation. For the benzene
molecule, these correlation are obtained by a coarse grain analysis in
which the ``site'' is defined to be a cylinder of radius $1.3~ a_0$ 
centered on the carbon nuclei, with a cut off core (i.e. we considered only
the points with $|z|> 0.8~ a_0$). All the results are pure expectation values
obtained from forward walking calculations.}
\end{figure}

Indeed, as shown in the inset of Fig.(\ref{spindimer}), 
 the dimer--dimer correlations 
of benzene are remarkably well reproduced  by the ones of 
the six site Heisenberg ring, whereas  the spin--spin correlation 
of the molecule appears to decay faster than the corresponding one of the 
 model.
 Though it is not possible to make conclusions 
on long range properties of  a finite molecular system, our results  
suggest that the benzene molecule can be considered closer to a spin 
liquid,  
 rather than to a dimerized state, because, as well known, the 
Heisenberg model ground state is a spin liquid and displays   
spontaneous dimerization only when a sizable 
 next-nearest frustrating superexchange interaction is turned
 on.\cite{affleck}   

As any meaningful variational ansatz, the RVB approach naturally brings a new 
way of understanding the many-body problem, as for instance the Hartree-Fock 
theory helped to interpret the periodic table of elements, 
or to establish on theoretical grounds the band theory of insulators. 
With the RVB paradigm, many unusual 
phenomena now appear to be possibly 
 explained in a simple and consistent framework: 
the role of correlation in Mott insulators, or the explanation of HTc
superconductivity, and finally the fractionalization of spin excitations, 
which was supposed to take place only in quasi-one dimensional systems, 
and instead it has been recently detected in higher dimensions.\cite{coldea}
All these phenomena cannot be understood not even qualitatively within a mean 
field Hartree--Fock theory, as the important ingredient missing in the 
latter approach is just the correlation, that can lead to essentialy new
effects.
 
In our opinion the RVB wave function is a natural extension of the 
Hartree-Fock one, to which it reduces whenever the correlation term is 
switched off. In some sense the determinantal part is useful to represent 
the electronic density and all the one body properties of an electronic system.
On the other hand the Jastrow term is necessary to take correctly 
into account the density--density correlation, $N(r)= < n_0 n_r >$.
The long range behavior of $N(r)$ discriminates a metal, displaying Friedel
oscillations at $4 k_F$ where $k_F$ is the Fermi momentum, from an insulator, 
which shows an exponentially localized correlation 
$N(r) \simeq \exp (-r /\xi) $, where $\xi$ is the corresponding 
characteristic length.
The Jastrow correlation can become non trivial when the 
determinantal part acquires a non conventional meaning. 
For instance, the determinantal part in the RVB wave function 
could describe a superconductor or a metal, 
but the presence of the Jastrow factor is able to turn the system into an
insulator, by correlating the electrons in a non trivial way.
On the other hand, superconductivity can naturally become 
stable in a system with only repulsive interactions, despite the BCS theory
would require an effective attraction mediated by the phonons. 

For all the above reasons we believe that it is the right time to make 
an effort to study complex electronic systems by means of this new paradigm, 
especially for discovering new challenging effects in which the role of 
correlation is dominant.

% Bibliography style file from Elsevier
\bibliographystyle{elsart-num}
\bibliography{rvbproceeding}

\end{document}